\documentclass{appolb}

\usepackage{amsfonts,amsmath,amssymb,bm}
\usepackage{graphicx}

\begin{document}
\title{ACCURACY ANALYSIS OF THE BOX-COUNTING ALGORITHM}
\author{A. Z. G\'orski$^1$, S. Dro\.zd\.z$^{1,2}$, A. Mokrzycka$^{1,3}$, J. Pawlik$^{1,3}$
\address{ %
$^1$H. Niewodnicza\'nski Institute of Nuclear Physics, Polish Academy of Sciences, Radzikowskiego 152, Krak\'ow, 31-342, Poland \\%
$^2$Faculty of Math. \& natural Sci., Univ. of Rzesz\'ow, 35-310 Rzesz\'ow, Poland \\%
$^3$AGH University of Science and Technology, Faculty of Physics and Applied Computer Science, Krak\'ow, Poland}}
\maketitle
\begin{abstract}
Accuracy of the box-counting algorithm for numerical computation of the fractal exponents 
is investigated. To this end several sample mathematical fractal sets are analyzed. 
It is shown that the standard deviation obtained for the fit of the fractal scaling 
in the log-log plot strongly underestimates the actual error. 
The real computational error was found to have power scaling with respect to the number of data 
points in the sample ($n_{tot}$). 
For fractals embedded in two-dimensional space the error is larger than for those embedded
in one-dimensional space. For fractal functions the error is even larger. 
Obtained formula can give more realistic estimates for the computed generalized 
fractal exponents' accuracy.
\end{abstract}


\section{Introduction}

 In last decades computations of fractal dimensions (exponents)
have become very popular in various areas of physics, as well 
as in interdisciplinary research. Fractal structures have been 
found in wide spectrum of problems, ranging from high energy physics 
\cite{Bialas1988} to cosmology \cite{Chmaj1991} and from 
medicine \cite{AZG-Skrzat} to econophysics \cite{AZG2002}. 
In spite of its popularity accuracy of obtained results is usually 
either not discussed or overestimated. 
Moreover, it has been found that in quite a few papers 
wrong numerical results and conclusions have been published
\cite{AZG-Skrzat,AZGpreprint,AZG2001,McCauley2002}. 

 The aim of this paper is to calculate fractal exponents for several well known 
mathematical fractals with the box-counting algorithm 
to estimate real accuracy of these computations. 
The dependence of accuracy with respect to the size of the available data 
set ($n_{tot}$) is also discussed and its simple scaling properties are found. 

 It should be stressed that accuracy of fractal exponent computations 
in principle depends on many factors. For example, the accuracy can be 
degraded by presence of noise in the data. Also, one can get different results
using different box-counting algorithms (see {\it e.g.} \cite{Theiler1990})
or using different digital representations of the investigated physical object
(picture). Furthermore, one should be very careful translating Hurst exponents
into fractal exponents as, in general, there is no simple reation of both \cite{Jaffard1997}. 
 
 In Sec.~2 we calculate the box-counting fractal exponents for six different
fractal sets for various numbers of data points ($n_{tot}$). The generalized fractal exponent 
is defined in the standard way \cite{AZG2001}
\begin{equation}
\label{dqdef}
  d(q) = \frac{1}{1-q} \, \lim_{N\to\infty} 
  \frac{\sum_i \log p_i^q(N)}{\log N}
 \ ,
\end{equation}
where $N$ denotes the total number of boxes and $p_i(N)$ is the measure of the subset
in the $i$-th box for the given division $N$. Where the box size $\varepsilon = 1/N$. 
To find accuracy estimates the obtained results are compared with precise mathematical values
of the exponents determined analytically. Furthermore, we calculate standard errors
for the linear fits in the log-log plots used to calculate the exponents. 
Finally, the inverse power fits were found to give fair approximation of accuracy dependence 
on the size of the data set ($n_{tot}$). 
The final Section contains summary and conclusions.

\section{Accuracy estimates}

To start with we calculate fractal exponents for fractal sets embedded in one-dimensional 
space, namely the classical Cantor set (CS) \cite{Mandelbrot} and the (multifractal) 
asymmetric Cantor set (ACS) \cite{Tel1987}. 
The calculations have been repeated for different sizes of the sets, ranging from
less than $10^2$ up to $10^5$ data points. 
The final results are given in Fig.~1. Crosses indicate the real accuracy of the box-countong 
algorithm computations, \textit{i.e.} the absolute value of the difference between
the calculated exponent and exact analytical result. Circles give standard errors
obtained for the linear fits in the corresponding log-log plots. In addition, the inverse power
fit for accuracy as the function of the number of available data points ($n_{tot}$) 
is given by the dashed line. The parameter $\alpha$ denotes the exponent of the 
inverse power fit
\begin{equation}
\label{alphadef}
  \text{real error} \ \sim \ \frac{1}{n_{tot}^\alpha}
 \ .
\end{equation}
 At first glimp it is clear that the standard error, that is often treated 
as the accuracy of the algorithm, considerably (up to the order of magnitude) 
underestimates the real error. In the case of Cantor set (A) we have 
$\alpha_{CS} \approx 0.50$. 
Similar result was obtained for the case of asymmetric Cantor set, 
$\alpha_{ACS} \approx 0.48$. Hence, in these cases one can expect the error 
of the size $\sim 1/\sqrt{n_{tot}}$.

\begin{figure}
\begin{center}
 \includegraphics[width=8.0cm,angle=0]{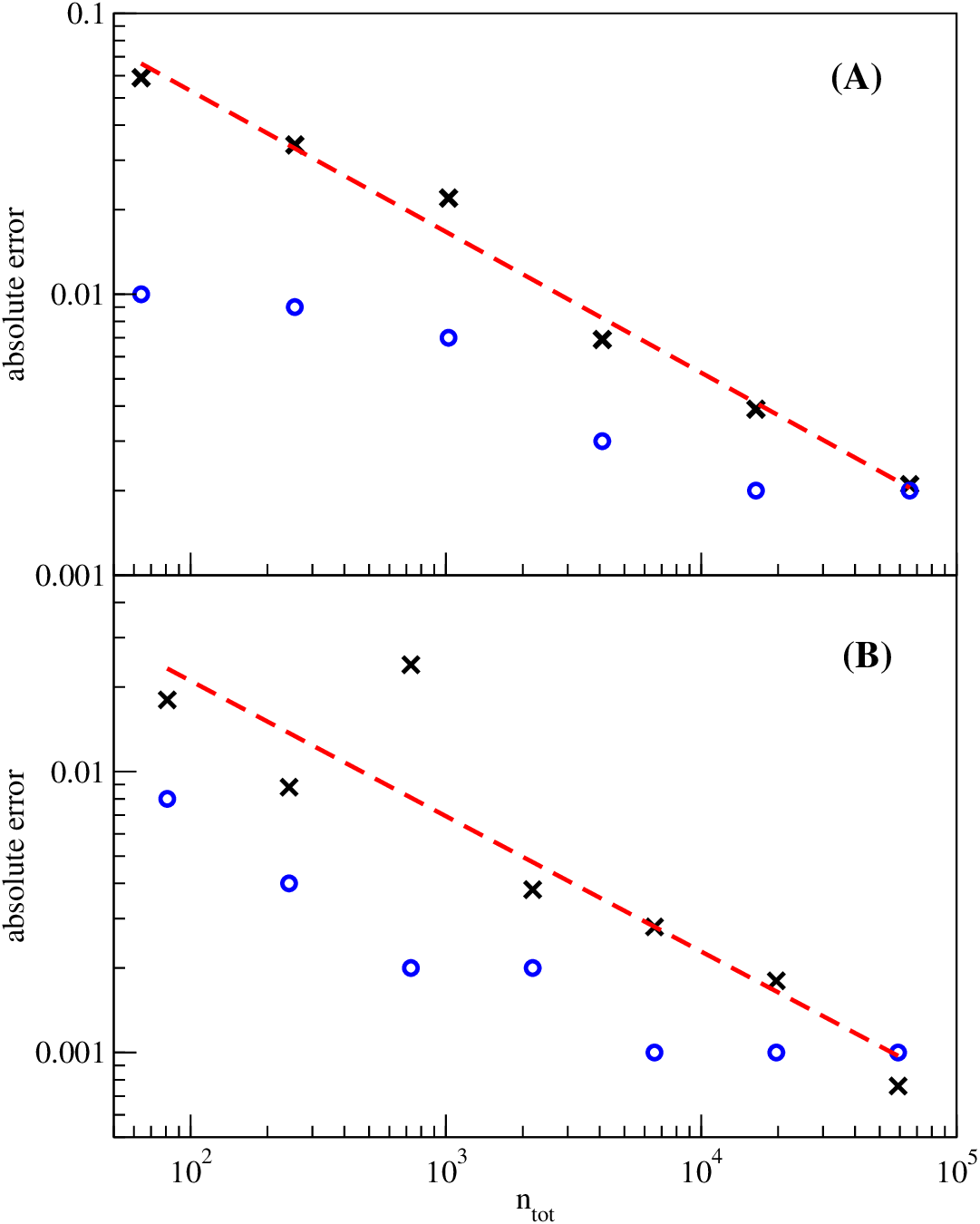}
 \caption{Real accuracy of the fractal exponent $d(0)$ (crosses) 
 and the standard errors obtained for the linear fit in the log-log plots
 used to determine fractal exponents (circles). The dashed line is the inverse power fit.
 The upper plot (A) is for the Cantor set and the lower plot (B) is for the ACS.}
\end{center}
\label{fig:fig1}
\end{figure}

\begin{figure}
\begin{center}
 \includegraphics[width=8.0cm,angle=0]{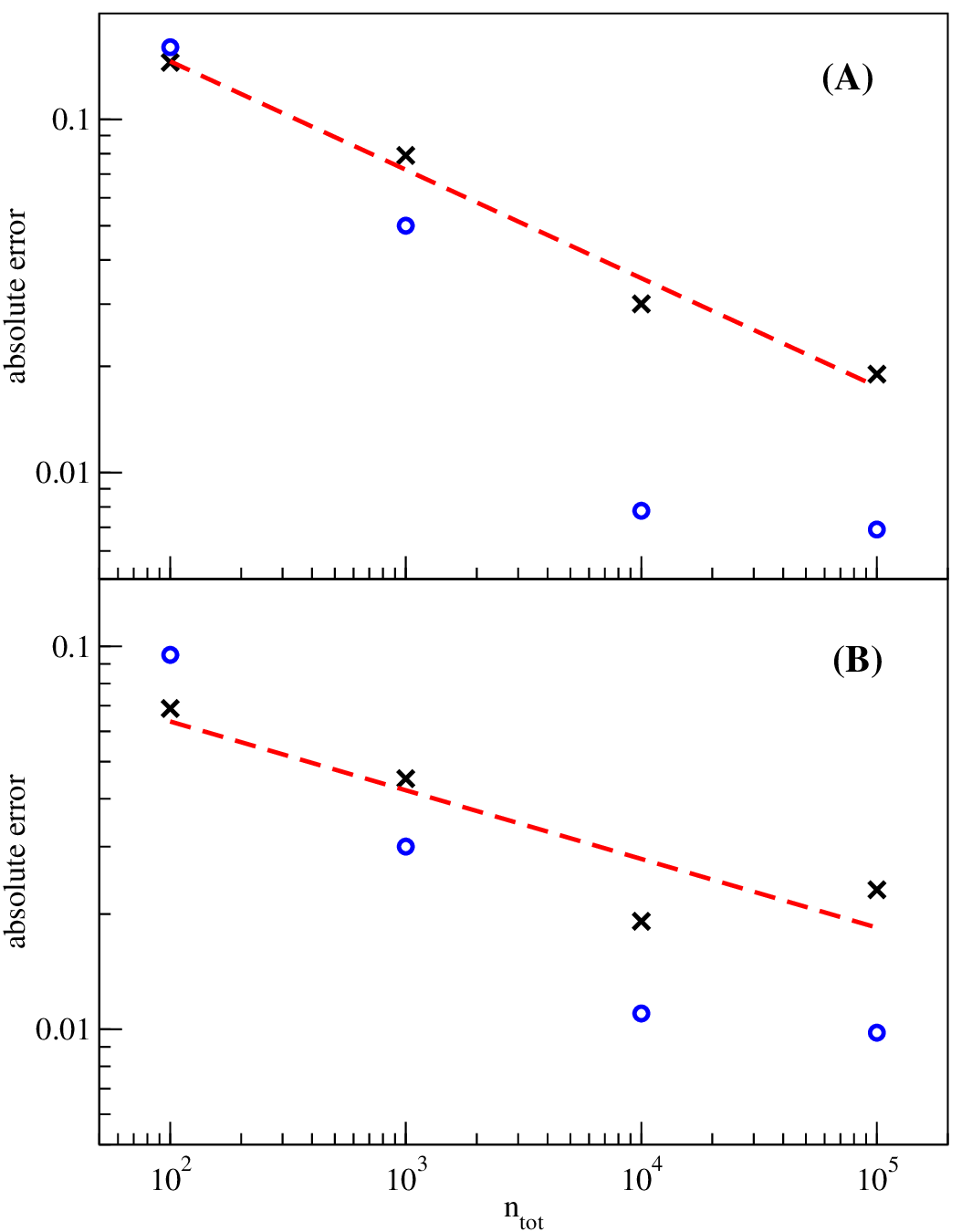}
 \caption{Real accuracy of the fractal exponent $d(0)$ (crosses) 
 and the standard errors obtained for the linear fit in the log-log plots
 used to determine fractal exponents (circles). The dashed line is the inverse power fit.
 The upper plot (A) is for the Sierpi\'nski triangle and the lower plot (B) 
 is for the Koch curve.}
\end{center}
\label{fig:fig2}
\end{figure}

\begin{figure}
\begin{center}
 \includegraphics[width=8.0cm,angle=0]{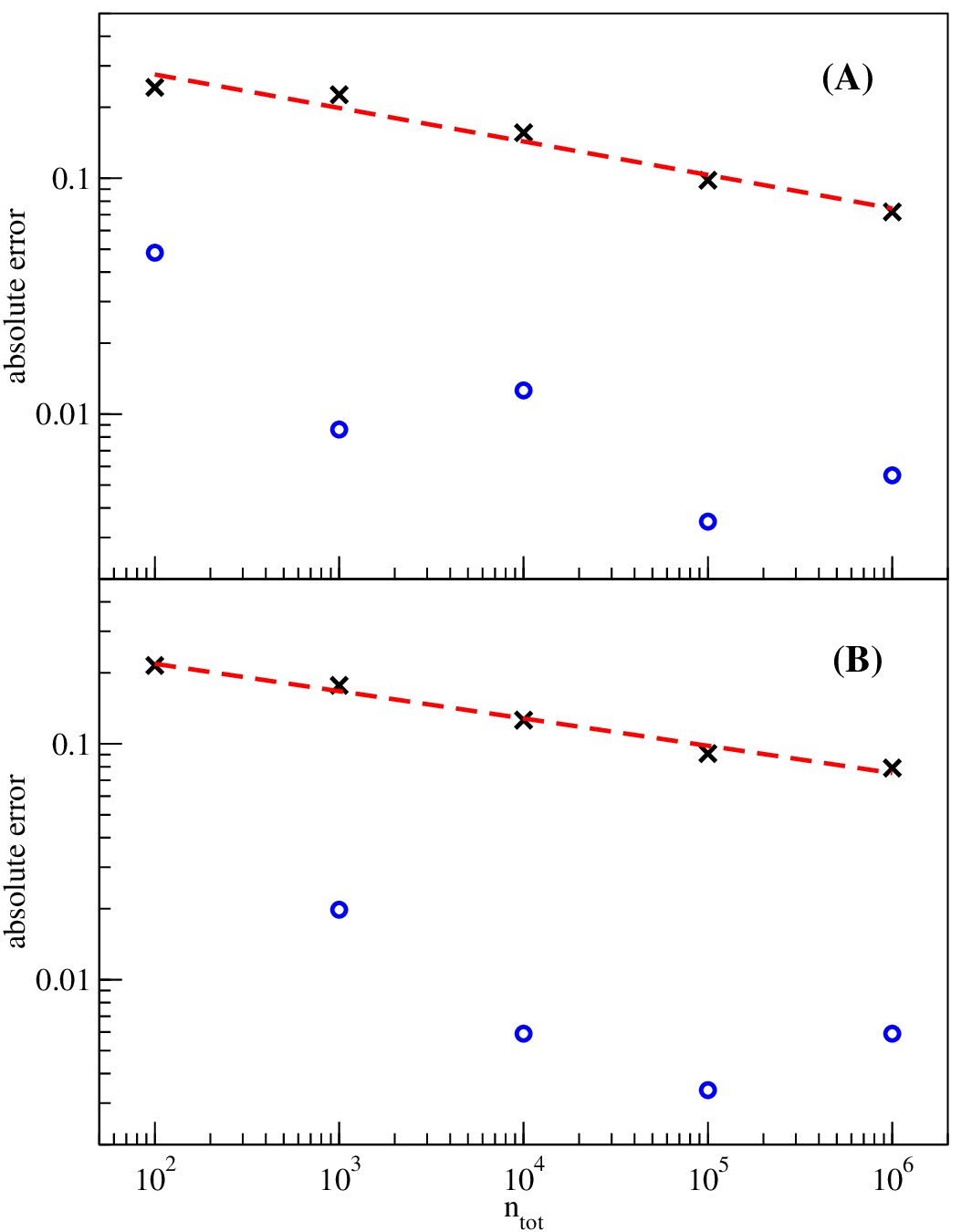}
 \caption{Real accuracy of the fractal exponent $d(0)$ (crosses) 
 and the standard errors obtained for the linear fit in the log-log plots
 used to determine fractal exponents (circles). The dashed line is the inverse power fit.
 The upper plot (A) is for the Weierstrass-Mandelbrot curve with parameter $D=1.5$ and 
 the lower plot (B) is for the curve with $D=1.8$.}
\end{center}
\label{fig:fig3}
\end{figure}

 As the second step we analyze fractal sets embedded in two-dimensional space:
the Sierpi\'nski triangle and the Koch curve. The results are given in Fig.~2,
with the same notation as for the Fig.~1. 
In this case we have $\alpha_{ST}\approx 0.31$ and $\alpha_{KC}\approx 0.18$.
Hence, the error scales approximately as $\sim 1/n_{tot}^{1/4}$. 
This results is intuitively clear, as to have the same accuracy as for the
one-dimensional case the squared number of data points has to be used. 
In these examples the actual error is also much bigger than the estimated 
standard error.

 For fractals embedded in two-dimensional space special attention should 
be paid to fractals that are of the function type, \textit{i.e.} there exists
a reference frame in which for a given value of one coordinate there is
at most one point of the fractal set. Hence, in one direction number of points
in a given box is limited and this may cause slower convergence of the box-counting 
scheme (scaling can be observed in one direction only) resulting in bigger errors, 
slower convergence. 
Because this type of fractals has wide applicability, \textit{e.g.} 
in the time series analysis, we will consider this case separately. 
A good example of such fractal sets with precisely known fractal dimensions are 
the Weierstrass-Mandelbrot (WM) functions \cite{BerryLewis}
\begin{equation}
\label{WMdef}
 W(t) = \sum_{n=-\infty}^{n=+\infty} \ \frac{1}{\gamma^{(2-D)n}} \ 
 \left[ 1 - \cos (\gamma^n t) \right]
 \ .
\end{equation}

 We investigate two WM fractal functions with dimensions $d=1.5$ (A) and $d=1.8$ (B). 
The results are shown in Fig.~3.  Again one can find a fair inverse power fit
for the error with the exponents $\alpha_{WM}$ equal to $0.14$ and $0.12$
($\sim 1/8$), respectively. 
Hence, to have similar accuracy as for ordinary fractals embedded in two-dimensional
space one has to use squared number of data points ($n_{tot}$). 
This is intuitively clear, as in this case the fractal scaling can be observed 
in only one (instead of two) dimensions. 
In effect, to have reasonable accuracy for the fractal exponent, a very large number 
of data points has to be taken into account ($>10^5$) even though we deal with perfect 
mathematical fractals without any external noise.

\section{Summary and conclusions}

 It has been shown that for the box-counting algorithm there is a fair inverse power
scaling of the actual error of the computed fractal exponents of the type (\ref{alphadef}).
The standard error calculated for the fit in the log-log plot strongly underestimates 
the actual error leading to the overestimated accuracy. 
Furthermore, to obtain a given level of accuracy number of data points ($n_{tot}$)
used for fractals embedded in two-dimensional space should be the squared number of 
data points sufficient for fractals embedded in one-dimensional space.
The corresponding exponents $\alpha$ are roughly equal to $1/4$ and $1/2$, respectively. 

 Similar phenomenon occurs for fractal functions embedded in two-dimensional space,
where the exponent $\alpha$ was found around $1/8$. Hence, up to an overall factor, 
to have the same accuracy as for the Cantor set and $10^2$ data points one has to use 
of order $10^8$ data points for the fractal WM function. 

 The formula (\ref{alphadef}) and plots in Figs.~1--3 can be used to estimate 
accuracy of such computationion in much higher accuracy than the estimated 
standard errors for the linear fits. 
The estimated accuracy of the box-counting algorithm for various sizes of the
data sets ($n_{tot}$) is also presented in Table~1.
It should be stressed that for fractals with additional external noise 
one can expect even worse results --- the errors will be greater.

\begin{table}
\begin{center}
 \caption{\label{tab:table1} Absolute errors. }
     \begin{tabular}{|lccc|}%
      \hline %
      \hline %
         $n_{tot}$ & $1000$ & $10~000$ & $100~000$\\
      \hline%
      \hline%
         1-D fractals   & $\pm$0.020 & $\pm$0.006 & $\pm$0.002 \\
         2-D fractals   & $\pm$0.060 & $\pm$0.030 & $\pm$0.020 \\
         2-D W-M curve  & $\pm$0.200 & $\pm$0.150 & $\pm$0.100 \\
     \hline %
     \hline %
\end{tabular}
\end{center}
\end{table}


\begin{thebibliography}{99}

\begin{small}

\bibitem{Bialas1988}
A. Bia\l as, R. Peszanski, Nucl. Phys. {\bf B 308}, 803 (1988). 

\bibitem{Chmaj1991}
T. Chmaj, W. Doroba, W. S\l omi\'nski, Z. Phys. {\bf C50}, 333 (1991).

\bibitem{AZG-Skrzat}
A. Z. G\'orski, J. Skrzat, 
J. Anat. {\bf 208}, 353 (2006).

\bibitem{AZG2002}
A.Z. G\'orski, S. Dro\.zd\.z, J. Speth, Physica {\bf A316}, 496 (2002).

\bibitem{AZGpreprint}
A. Z. G\'orski,
{\it Comment on fractality of quantum mechanical energy spectra},
preprint arXiv chao-dyn/9804034.

\bibitem{AZG2001}
A. Z. G\'orski,
J. Phys. {\bf A34}, 7933 (2001).

\bibitem{McCauley2002}
J. L. McCauley, 
Physica {\bf A309}, 183 (2002).

\bibitem{Theiler1990}
J. Theiler, J. Opt. Soc. Am {\bf A7}, 1055 (1990).

\bibitem{Jaffard1997}
S. Jaffard, 
SIAM J. Math. Anal. {\bf 28} 944; 971, (1997).

\bibitem{Mandelbrot}
B. B. Mandelbrot, \textit{Fractals: Form, Chance, and Dimension} 
(San Francisco: Freeman 1977).

\bibitem{Tel1987}
T. Tel, T. Vicsek,
J. Phys. {\bf A20}, L835 (1987).

\bibitem{BerryLewis}
M. V. Berry, Z. V. Lewis,
Proc. R. Soc. Lond. {\bf A370}, 459 (1980).

\end{small}

\end{thebibliography}
\end{document}